\documentclass[aps,prl,preprint]{revtex4}

\usepackage{epsfig}

\draft

\begin{document}
\author{Hong-shi Zong$^{1,2}$, Jia-lun Ping$^{3}$, Hong-ting Yang$^{1}$,
Xiao-fu L\"{u}$^{2,4}$, and Fan Wang$^{1}$}
\address {$^{1}$ Department of Physics, Nanjing University, Nanjing 210093, P. R. China}
\address{$^{2}$ CCAST(World Laboratory), P.O. Box 8730, Beijing 100080, P. R. China}
\address{$^{3}$ Department of physics, Nanjing Normal University, Nanjing 210097, P. R. China}
\address{$^{4}$  Department of Physics, Sichuan University,
Chengdu 610064, P. R. China}

\title{The Calculation of Vacuum Properties 
from the Global Color Symmetry Model}

\begin{abstract}
A modified method for calculating the non-perturbative quark vacuum condensates from the global color symmetry model
is derived. Within this approach it is shown that the vacuum condensates are free of ultraviolet divergence which is
different from the previous studies. As a special case the two quark condensate $\langle\bar{q}q\rangle$ and the mixed quark gluon condensate $g\langle\bar{q}G_{\mu\nu}\sigma^{\mu\nu}q\rangle$ are calculated.
A comparison with the results of other nonperturbative QCD approaches is given.

\bigskip

Key-words: Non-perturbative methods in QCD, GCM, Vacuum condensate.

\bigskip

E-mail: zonghs@chenwang.nju.edu.cn.

\bigskip

\pacs{PACS Numbers: 24.85.+p, 12.38.Lg, 12.38.-t, 11.15Pg}

\end{abstract}

\maketitle

\begin{center}
{\bf \large I. INTRODUCTION}
\end{center}

The non-perturbative structure of the QCD vacuum is characterized by
various condensates: such as the two quark condensate $\langle\bar{q}q\rangle$, the mixed
quark-gluon condensate $g\langle\bar{q}G_{\mu\nu}\sigma^{\mu\nu}q\rangle$, the four
quark condensate $\langle\bar{q} \Gamma q\bar{q} \Gamma q\rangle$, etc. These condensates are essential for describing the physics
of the strong interaction[1,2]. There has been growing interest in describing the properties of hadrons in nuclear matter in terms of the in-medium quark and gluon condensates, which are shifted from their vacuum values[3-8]. Since the Global Color Symmetry Model(GCM)[9] provides a nonperturbative framework which admits a simultaneous study of the spontaneous chiral symmetry  breaking and confinement, it is useful in exploring the phase transition at nonzero temperature and density. Before we generalize GCM from zero temperature and density to nonzero temperature and density, it is necessary to check whether GCM can provide a good description of vacuum properties at zero temperature and density.

It is the aim of this paper to consider quark vacuum condensates in general and in particular the two quark condensate and the mixed quark-gluon condensate in the framework of the global color symmetry model. Although GCM violates local color $SU(3)_{C}$ gauge invariance and renormalizability, it provides a very successful description of various nonperturbative aspects of strong interaction physics and hadronic phenomena at low energies. These include, for instance, quark confinement[10], $U_{A}(1)$ breaking, and the $\eta-\eta'$ mass splitting[11], low energy chiral dynamics of Goldstone bosons$(\pi,K,\eta)$[12-18], meson form factors[19], heavy-light mesons[20], systems at finite temperate[21], or soliton[22] and Fadeev[23] descriptions of the nucleon. The first estimations of the two quark and the mixed quark-gluon condensates in GCM was made in Ref.[24], where the two quark condensate $\langle\bar{q}q\rangle$ itself is systematically smaller than the ``standard'' value of $-(250 MeV)^3$[1] 

Here, we present a modified method for calculating the quark vacuum condensate in the framework of GCM. Since the $u$, $d$ current quark mass is small, we consider the GCM generating functional for massless quarks, i.e., the chiral limit, in Euclidean space

\begin{equation}
Z[\bar{\eta}, \eta] = \int {\cal{D}}\bar{q} {\cal{D}} q {\cal{D}} A\exp\left\{-S_{GCM}+ \int d^4 x (\bar{\eta} q + \bar{q} \eta)\right\} \, ,
\end{equation}
where
$$ S_{GCM}=\int d^4 x~ \bar{q}(x)\left[\gamma\cdot\partial
-ig\frac{\lambda^{a}_{C}}{2}\gamma\cdot A^{a}(x)\right]q(x) +\int d^4 x d^4 y~ 
\frac{1}{2}A^a_{\mu}(x)\left[D^{ab}_{\mu\nu}(x-y)\right]^{-1}A^b_{\nu}(y), $$
with $D^{ab}_{\mu\nu}(x-y)$ denoting the gluon two-point green function. Because the form of $D^{ab}_{\mu\nu}(x-y)$ in the infrared region is unknown, one often treats the $D^{ab}_{\mu\nu}(x-y)$ as the GCM input parameter, which, as we will discuss later, is chosen to reproduce certain aspects of low energy hadronic properties.  For convenience we will use the Feynman like gauge $ D_{\mu\nu}^{ab} (x-y)=\delta_{\mu\nu}\delta^{ab} D(x-y)$ from now on.

Performing the functional integral over ${\cal{D}} A$ in Eq.(1)(see Eq.(27) below),  
we obtain the GCM generating functional
\begin{equation}
Z[\bar{\eta},\eta] = \int {\cal{D}}\bar{q} {\cal{D}} q ~e^{\left\{-\int d^4 x~ \bar{q}(x) \gamma\cdot\partial q(x)-\int d^4 x d^4 y~\frac{g^2}{2}j^a_{\mu}(x)D(x-y)j^a_{\mu}(y)+\int d^4 x~(\bar{\eta} q + \bar{q} \eta)\right\}}, 
\end{equation}
here  $j^a_{\mu}(x)$ denotes the color octet vector current $ j^a_{\mu}(x)=\bar{q}(x)\gamma_{\mu}\frac{\lambda^a_{C}}{2}q(x)$.

Introducing an auxiliary bilocal field $B^{\theta}(x,y)$ as in Ref.[9],
the generating functional of GCM can be written as
\begin{eqnarray}
Z[\bar{\eta}, \eta] & = &
\int {\cal{D}}\bar{q} {\cal{D}} q {\cal{D}} B^{\theta}(x,y)
\exp\left\{-S [ \bar{q}, q, B^{\theta}(x,y)] + \int d^4 x ( \bar{\eta} q +
\bar{q} \eta)\right\} \, ,
\end{eqnarray}
where
$$ S[\bar{q}, q, B^{\theta}(x,y)] = \int \int d^4 x d^4 y \left[ \bar{q}(x)
{\cal{G}}^{-1}[x,y;[B^{\theta}]] q(y) + \frac{B^{\theta}(x,y) B^{\theta}(y,x)}
{2 g^2 D(x-y)} \right] \; , $$
with
\begin{equation}
{\cal{G}}^{-1}[x,y; [B^{\theta}]] = \gamma \cdot \partial_{x} \delta^{(4)}(x-y) +
\frac{1}{2} \Lambda^{\theta} B^{\theta}(x,y) \, ,
\end{equation}
where the matrices
$\Lambda^{\theta} = D^{a} F^{b} C^{c}$ is determined by Fierz
transformation in Dirac, flavor and color space, and are given by
\[
\Lambda^{\theta}=\frac{1}{2} (1_{D},i\gamma_{5},\frac{i}{\sqrt{2}}\gamma_{\mu},\frac{i}{\sqrt{2}}\gamma_{\mu}\gamma_{5})\otimes(\frac{1}{\sqrt{3}}1_{F},\frac{1}{\sqrt{2}}\lambda^{a}_{F})\otimes(\frac{4}{3}1_{C},\frac{i}{\sqrt{3}}\lambda^{a}_{C}).
\]

Performing the functional integral over ${\cal{D}} \bar{q}$ and
${\cal{D}} q$ in Eq.(3), we obtain the GCM generating functional
\begin{equation}
Z[\bar{\eta}, \eta] = \int {\cal{D}} B^{\theta}(x,y) \exp(-S [ \bar{\eta},
\eta, B^{\theta}(x,y)] ) \, ,
\end{equation}
where
\begin{eqnarray}
S[\bar{\eta}, \eta, B^{\theta}(x,y)] & = & - \mbox{Tr} \ln \left[\rlap/
\partial \delta (x-y) + \frac{1}{2} \Lambda^{\theta} B^{\theta}(x,y) \right]
\nonumber \\
& & + \int\int d^4 x d^4 y\left[\frac{B^{\theta}(x,y)
B^{\theta}(y,x)}{2 g^2 D(x-y)}
+ \bar{\eta}(x) {\cal{G}}(x,y;[B^{\theta}])\eta(y) \right] .
\end{eqnarray}

The saddle-point of the action is defined as
$\left.\delta S[\bar{\eta}, \eta, B^{\theta}(x,y)]/\delta B^{\theta}(x,y)
\right\vert _{\eta =\bar{\eta}
= 0} =0$ and is given by
\begin{equation}
B^{\theta}_{0}(x-y)=g^2 D(x-y) tr[\Lambda^{\theta} {\cal{G}}_{0}(x-y)],
\end{equation}
where ${\cal{G}}_{0}$ stands for ${\cal{G}}[B^{\theta}_{0}]$ and the 
trace in Eq.(7) is to be taken in Dirac and color space, whereas the flavor trace has been separated out.

We will calculate the vacuum condensates from the saddle-point expansion, that is , we will work at the mean field level. This is consistent with the large $N_{C}$ limit in the quark fields for a given model gluon two-point function. In the mean field approximation, the field $B^{\theta}(x-y)$ is substituted by their vacuum $B^{\theta}_{0}(x-y)$. Under this approximation, the dressed quark propagator $G(x-y)\equiv {\cal{G}}_{0}(x-y)$ in GCM has the decomposition
\begin{equation}
G^{-1}(p)\equiv i\gamma\cdot p+\Sigma(p)\equiv i\gamma\cdot p A(p^2)+B(p^2)
\end{equation}
with the self-energy dressing of the quarks $\Sigma(p)$ is defined as:
\begin{equation}
\Sigma(p)\equiv \frac{1}{2} \Lambda^{\theta}B^{\theta}_{0}(p)=\int d^4x e^{ip\cdot x}\left[\frac{1}{2} \Lambda^{\theta}B^{\theta}_{0}(x)\right] =i\gamma\cdot
p[A(p^2)-1]+B(p^2) ,
\end{equation}
where the self energy functions $A(p^2)$ and $B(p^2)$ are
determined by the rainbow Dyson-Schwinger equation[9]
\begin{equation}
[A(p^2)-1]p^2=\frac{8}{3}\int \frac{d^{4}q}{(2\pi)^4}g^2 D(p-q)
\frac{A(q^2)p\cdot q}{q^2A^2(q^2)+B^2(q^2)},
\end{equation}
\begin{equation}
B(p^2)=\frac{16}{3}\int \frac{d^{4}q}{(2\pi)^4}g^2 D(p-q)
\frac{B(q^2)}{q^2A^2(q^2)+B^2(q^2)}.
\end{equation}
In order to get the numerical solution of $A(p^2)$ and $B(p^2)$, one often uses model forms for the gluon two-point function as input in Eqs.(10-11). Here we investigate two different two parameter models for gluon propagator
\begin{equation}
g^2D^{(1)}(q^2)=g^2D^{(1)}_{IR}(q^2)+g^2D_{UV}(q^2)=3\pi^{2}\frac{\chi^2}{\Delta^2}e^{-\frac{q^2}{\Delta}}+\frac{4\pi^2 d}{q^2ln\left(\frac{q^2}{\Lambda^2_{QCD}}+e\right)},
\end{equation}
and
\begin{equation}
g^2D^{(2)}(q^2)=g^2D^{(2)}_{IR}(q^2)+g^2D_{UV}(q^2)=4\pi^2 d\frac{\chi^2}{q^4+\Delta}+\frac{4\pi^2 d}{q^2ln\left(\frac{q^2}{\Lambda^2_{QCD}}+e\right)}.
\end{equation}

The term $D_{IR}(q^2)$, which dominates for small $q^2$, simulates the infrared enhancement and confinement. The other term $D_{UV}(q^2)$ dominating for large $q^2$ is an asymptotic ultraviolet(UV) tail which matches the known one-loop renormalization group result with $d=[12/(33-2N_{f})]=12/27$, $\Lambda_{QCD}=200~MeV$. The model parameters $\chi$ and $\Delta$ are adjusted to reproduce the weak decay constant in the chiral limit $f_{\pi}=87~MeV$. The forms of $g^2D(q^2)$ have been used in Ref.[15] and it has been shown that with these values a satisfactory description of all low energy chiral observables can be achieved(more detail can be seen in Refs.[15] and [24]).

Here we want to stress that the $B(p^2)$ in Eqs.(10,11) has two qualitatively distinct solutions. The ``Nambu-Goldstone'' solution, for which
\begin{equation}
B(p^2)\neq 0,
\end{equation}
describes a phase in which: 1) chiral symmetry is dynamically broken. Because one has a nonzero quark mass function; and 2) the dressed quarks are confined, because the propagator described by these functions does not have a Lehmann representation[10]. In ``Nambu-Goldstone'' phase, the vacuum configuration $ B^{\theta}_{0}(x-y)$ in GCM(at the mean field approximation) can be regarded as a good approximation to the ``exact'' vacuum in QCD. The alternative ``Wigner'' solution, for which
\begin{equation}
B(p^2)\equiv 0,
\end{equation}
describes a phase in which chiral symmetry is not broken and the dressed-quarks are not confined. In ``Wigner'' phase, the vacuum configuration $ B^{\theta}_{0}(x-y)$ in GCM(at the mean field approximation) corresponds to the ``perturbative'' vacuum in QCD.

With these two ``phase'' characterized by qualitatively different quark propagators, the GCM can be used to calculate the vacuum condensates as we will show later.

Our paper is organized as follows: In Sec.II we briefly review the definition of vacuum condensate in QCD sum rule, and exhibit a modified method, which is consistent with this definition, for calculating the non-perturbative quark vacuum condensates from the GCM. Based on the modified approach, the two quark 
condensate and the mixed quark gluon condensate are calculated at the mean field level. 
A brief discussion and conclusion is given in Sec.III.

\begin{center}
{\bf \large II. EVALUATION OF VACUUM PROPERTIES}
\end{center}

In order to ensure that the treatment about the vacuum condensates in this paper is consistent with that in QCD sum rule, a brief introduction of vacuum condensates in QCD sum rule is described below:

In QCD sum rule, one often postulates that quark propagators are modified by the long-range confinement part of QCD; but the modification is soft in the sense that at short distance the difference between exact and perturbative
propagators vanishes.

To formalize this statement, one can write the ``exact'' propagator $G(x)$ as a vacuum expectation of a T-product of fields in the ``exact'' vacuum $|\tilde{0}\rangle$:
\begin{equation}
G_{ij}(x,y)\equiv\langle\tilde{0}|T[q_i(x)\bar{q}_j(y)]|\tilde{0}\rangle.
\end{equation}

According to the Wick theorem, one can write the T-product as the sum
\begin{equation}
T[q_i(x)\bar{q}_j(y)]=\underbrace{q_i(x)\bar{q}_j(y)}+:q_i(x)\bar{q}_j(y):
\end{equation}
of the ``pairing'' and the ``normal'' product. The ``pairing'' is just the expectation value of the T-product over the perturbative vacuum $|0\rangle$
\begin{equation}
\underbrace{q_i(x)\bar{q}_j(y)}=\langle 0|T[q_i(x)\bar{q}_j(y)]|0\rangle\equiv G_{ij}^{pert}(x,y),
\end{equation}
i.e., the perturbative propagator. By this definition, we have the following two quark vacuum condensate $\langle\tilde{0}|:\bar{q}(x)q(y):|\tilde{0}\rangle$:
\begin{eqnarray}
&&\langle \tilde{0}|:\bar{q}_i(x)q_j(y):|\tilde{0}\rangle=
\langle\tilde{0}|T[\bar{q}_i(x)q_j(y)]|\tilde{0}\rangle
-\langle 0|T[\bar{q}_i(x)q_j(y)]|0\rangle \\
&&=(-)\left[G_{ji}(y,x)-G_{ji}^{pert}(y,x)\right]\equiv (-)\Sigma_{ji}(y,x)
=(-)\int\frac{d^4 q}{(2\pi)^4}e^{i q\cdot (y-x)}\left[G(q^2)-G^{pert}(q^2)\right]_{ji},
\nonumber
\end{eqnarray}
where $\Sigma(x,y)=G(x,y)-G^{pert}(x,y)$. Thus, our assumption of $\langle\tilde{0}|:\bar{q}q:|\tilde{0}\rangle$ $\not=0$ is equivalent to the statement $G(x)\not= G^{pert}(x)$. Because at large momentum region the difference between exact and perturbative quark propagator vanishes, the quark condensate $\langle\tilde{0}|:\bar{q}q:|\tilde{0}\rangle$ is free of UV divergence. This conclusion is reasonable because the vacuum condensates reflects the IR behavior of QCD.

Similarly, we have the four quark vacuum condensate:
\begin{eqnarray}
&& \langle \tilde{0}|:\bar{q}(x)\Lambda^{(1)}q(x)\bar{q}(y)\Lambda^{(2)}q(y):|\tilde{0}\rangle \nonumber\\
&&=\langle \tilde{0}|T\left[\bar{q}(x)\Lambda^{(1)}q(x)
\bar{q}(y)\Lambda^{(2)}q(y)\right]|\tilde{0}\rangle
-\langle 0|T\left[\bar{q}(x)\Lambda^{(1)}q(x)
\bar{q}(y)\Lambda^{(2)}q(y)\right]|0\rangle \nonumber\\
&&-\langle \tilde{0}|:\bar{q}(x)\Lambda^{(1)}\underbrace{q(x)
\bar{q}(y)}\Lambda^{(2)}q(y):|\tilde{0}\rangle
-\langle \tilde{0}|:\underbrace{\bar{q}(x)\Lambda^{(1)}q(x)
\bar{q}(y)\Lambda^{(2)}q(y)}:|\tilde{0}\rangle\\
&&-\langle \tilde{0}|:\underbrace{\bar{q}(x)\Lambda^{(1)}q(x)}
\bar{q}(y)\Lambda^{(2)}q(y):|\tilde{0}\rangle
-\langle \tilde{0}|:\bar{q}(x)\Lambda^{(1)}q(x)
\underbrace{\bar{q}(y)\Lambda^{(2)}q(y)}:|\tilde{0}\rangle\nonumber,
\end{eqnarray}
here the $\Lambda^{(i)}$ stands for an operator in Dirac and color space. It should be noted that the treatment of vacuum condensates here is different from that in Refs.[24,25], where the contribution of the first term
of the right hand of Eq.(20) is considered to be the four quark vacuum condensate. By this definition, the vacuum condensates in Refs.[24,25] are just the corresponding quark green function taken at one point. This is not consistent with the definition of quark vacuum condensate in QCD sum rule.

In addition, we have the six quark vacuum condensate:
\begin{eqnarray}
&& \langle \tilde{0}|:\bar{q}(x)\Lambda^{(1)}q(x)\bar{q}(y)\Lambda^{(2)}q(y)
\bar{q}(z)\Lambda^{(3)}q(z):|\tilde{0}\rangle \nonumber\\
&&=tr[\Sigma(y,y)\Lambda^{(2)}]tr[\Sigma(z,x)\Lambda^{(1)}
\Sigma(x,z)\Lambda^{(3)}]
-tr[\Sigma(z,x)\Lambda^{(1)}\Sigma(x,y)\Lambda^{(2)}\Sigma(y,z)\Lambda^{(3)}]\nonumber\\
&&-tr[\Sigma(y,x)\Lambda^{(1)}\Sigma(x,z)\Lambda^{(3)}\Sigma(x,y)\Lambda^{(2)}]+tr[\Sigma(x,x)\Lambda^{(1)}]tr[\Sigma(x,y)\Lambda^{(2)}\Sigma(y,z)\Lambda^{(3)}]\\
&&+tr[\Sigma(z,z)\Lambda^{(3)}]tr[\Sigma(y,x)\Lambda^{(1)}
\Sigma(x,y)\Lambda^{(2)}]-tr[\Sigma(x,x)\Lambda^{(1)}]tr[\Sigma(y,y)\Lambda^{(2)}]tr[\Sigma(z,z)\Lambda^{(3)}]\nonumber.
\end{eqnarray}
Based on the above statement, in order to calculate quark vacuum condensates one must know not only the ``exact'' but also the ``perturbative'' quark propagator in advance. The calculation of the ``exact'' quark propagator in GCM(at the mean field approximation) has been given by Eqs.(10,11) and Eq.(14), the left question now is how to treat consistently the ``perturbative'' quark propagator in GCM.

In ``Wigner'' phase(the perturbative quark scalar self energy function
$B'(p^2)\equiv 0$), the Dyson-Schwinger equation(10,11) reduces to:
\begin{equation}
[A'(p^2)-1]p^2=\frac{8}{3}\int \frac{d^{4}q}{(2\pi)^4}g^2 D(p-q)
\frac{p\cdot q}{q^2A'(q^2)},
\end{equation}
where $A'(p^2)$ denotes the perturbative quark vector self energy function. Therefore, the perturbative quark propagator in GCM can be written as $G^{pert}(q^2)=\frac{-i\gamma\cdot q}{A'(q^2)q^2}=-i\gamma\cdot q C(q^2)$.  Numerical studies show that for $q^2\gg \Lambda^2_{QCD}$, one has
\begin{equation}
G(q^2)-G^{pert}(q^2)=\frac{-i\gamma\cdot q A(q^2)+B(q^2)}{A^2(q^2)q^2+B^2(q^2)}+i\gamma\cdot q C(q^2)=0,
\end{equation}
which is just what one would expect in advance. The above results can be seen from the following figures (fig.1--fig.4).

\epsfxsize=3.0in \epsfbox{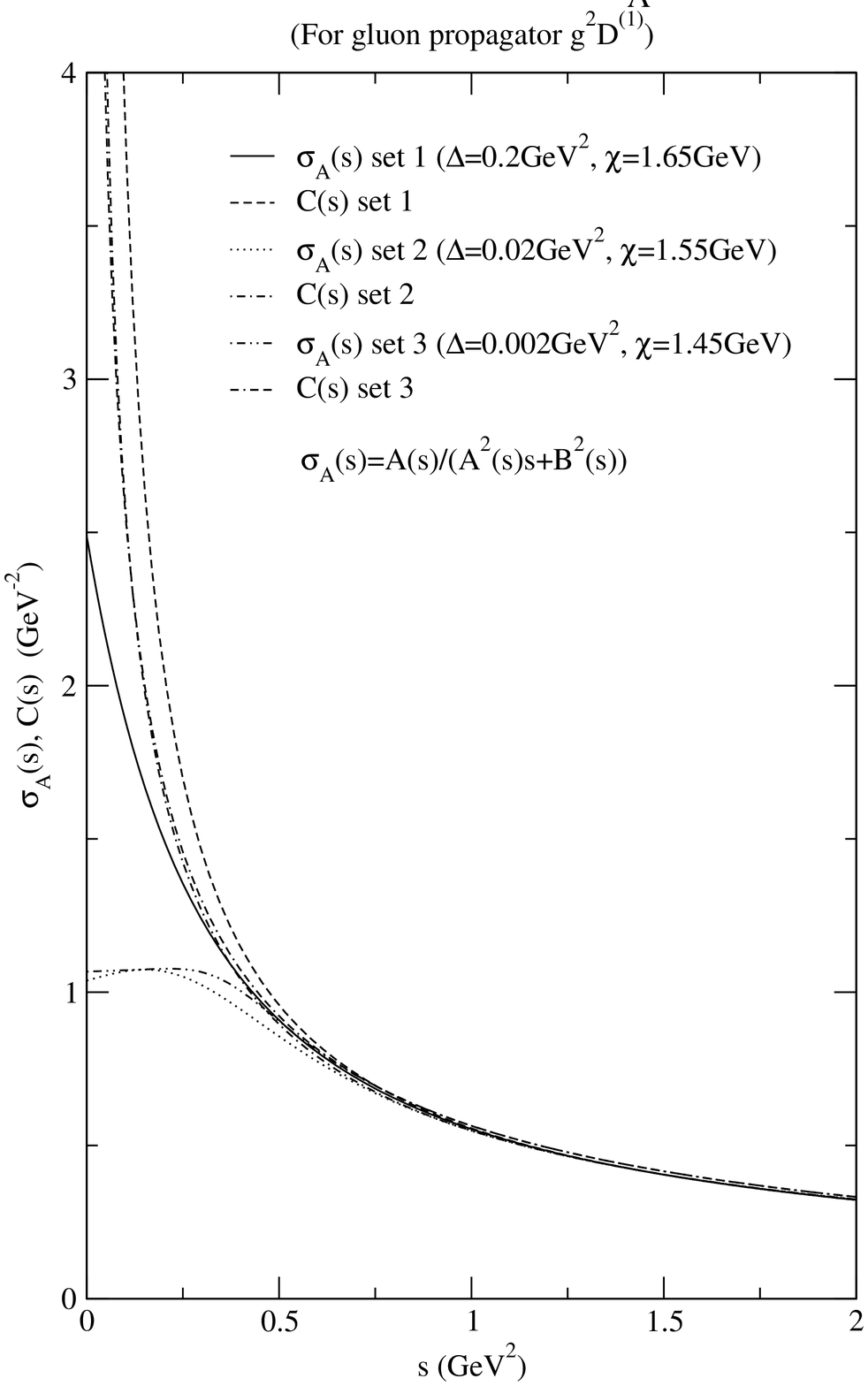}
\epsfxsize=3.0in \epsfbox{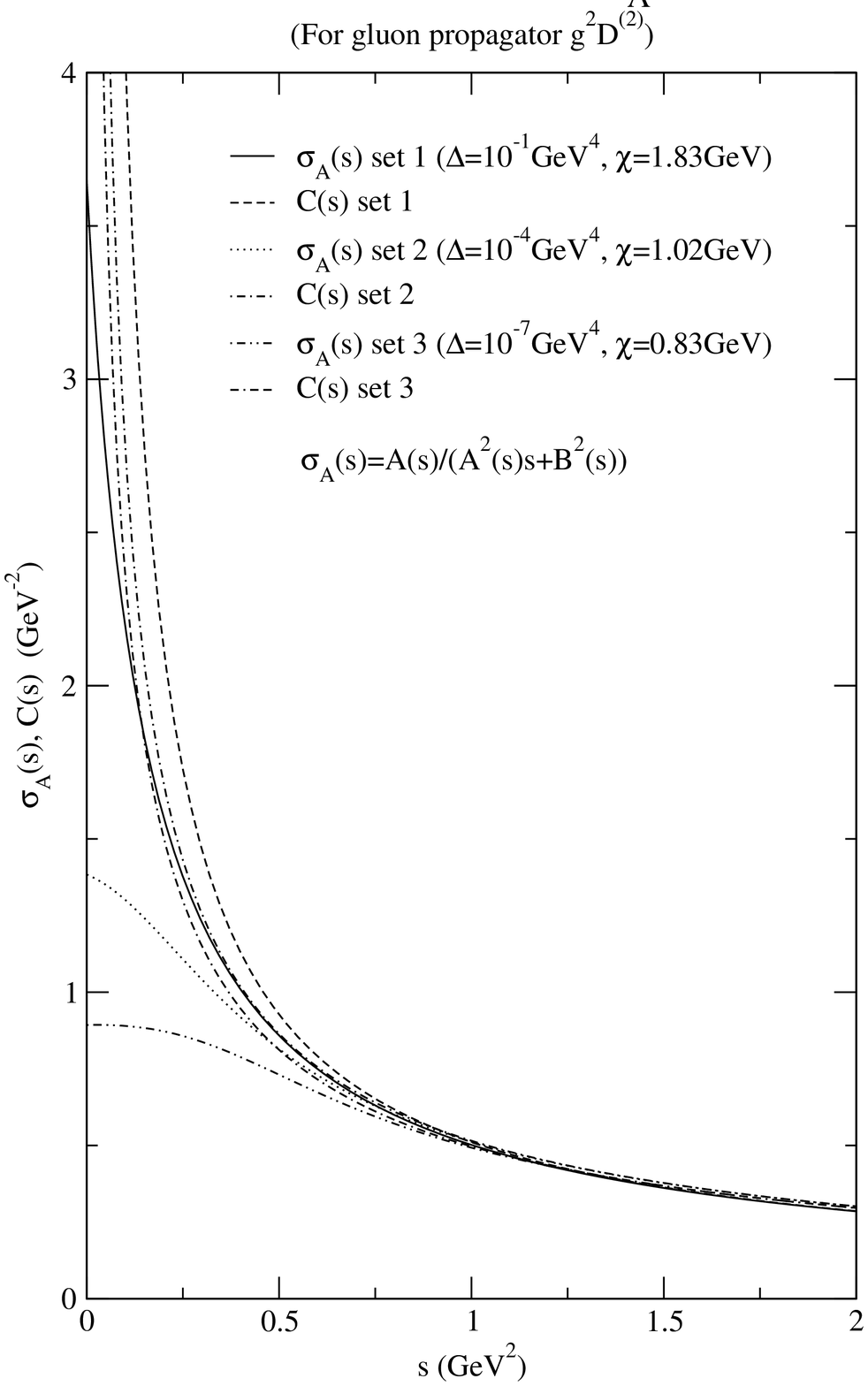}

\epsfxsize=3.0in \epsfbox{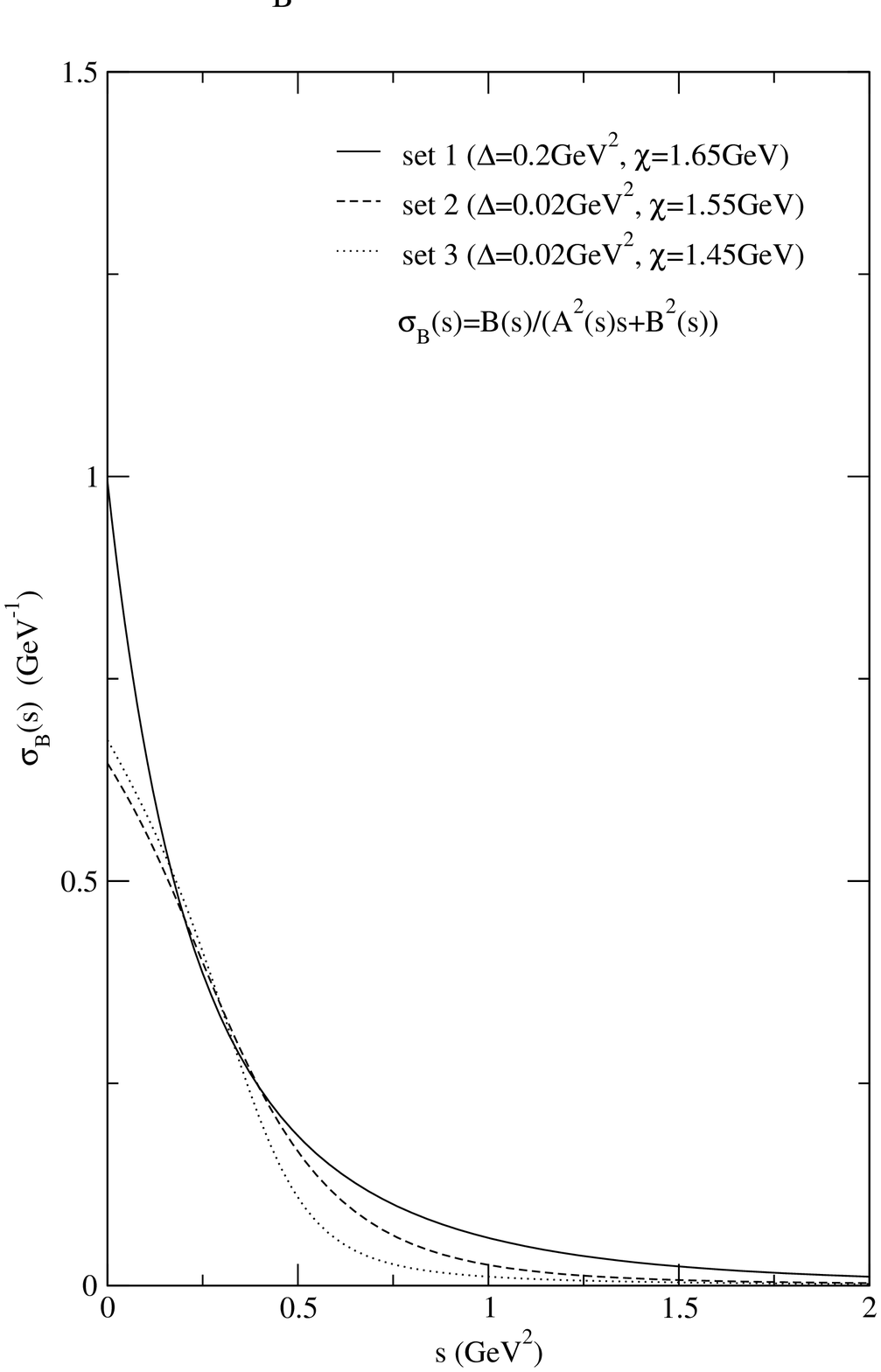}
\epsfxsize=3.0in \epsfbox{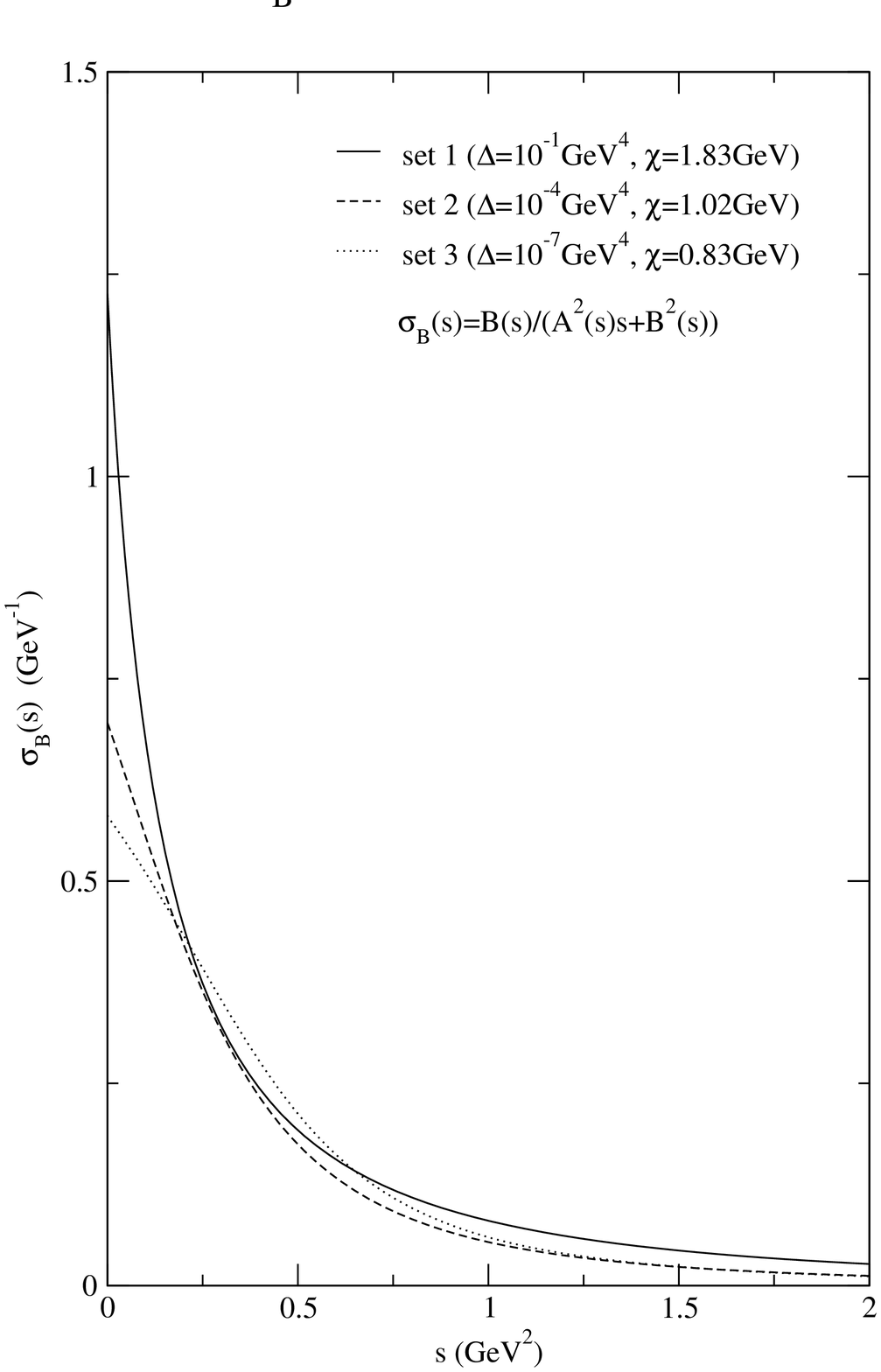}

Once the ``exact'' and ``perturbative'' quark propagator in our model are determined,
one can calculate the $\langle\tilde{0}|:\bar{q}q:|\tilde{0}\rangle$,
$\langle\tilde{0}|:\bar{q} \Lambda^{(1)} q\bar{q} \Lambda^{(2)} q:|\tilde{0}\rangle$ vacuum condensate and the mixed quark gluon condensate $g\langle\bar{q}G_{\mu\nu}\sigma^{\mu\nu}q\rangle$ at
the mean field level. In particular we obtain the two quark condensate $\langle\tilde{0}| :\bar{q}q:|\tilde{0}\rangle$ in the chiral limit;
\begin{eqnarray}
&&\langle\tilde{0}| :\bar{q}q:|\tilde{0}\rangle
=(-)\lim_{x\to0}{tr[\Sigma(x,0)]}\nonumber\\ 
&=&(-)\lim_{x\to0}{tr[G(x,0)]}=(-)\lim_{x\to0}\int\frac{d p^4}{(2\pi)^4}e^{i p\cdot x}
\frac{12~B(p^2)}{p^2A^2(p^2)+B(p^2)}\nonumber\\ 
&=&(-)\int\frac{d p^4}{(2\pi)^4}\frac{12~B(p^2)}{p^2A^2(p^2)+B(p^2)}=(-)\frac{3}{4\pi^2}\int^{\infty}_{0} ds \frac{sB(s)}{sA^2(s)+B(s)},
\end{eqnarray}
where we have used the fact that $tr[G^{pert}(x,0)]=0$ in the chiral limit. Due to the contribution of the subtraction term vanishes, this is the quark Green function taken at one point, which in momentum space is a closed quark loop. Therefore, in principle, the upper limit of integration in Eq.(24) should be taken to be infinity. 

It should be noted that the upper limit of integration in Eq.(24) is different from that in Refs.[24,25]. In Refs.[24,25], the two quark condensate is calculated as
\[
\langle\tilde{0}| :\bar{q}q:|\tilde{0}\rangle_{\mu}^{Meissner}
=(-)\frac{3}{4\pi^2}\int^{\mu=1 GeV^2}_{0} ds s
\frac{B(s)}{sA^2(s)+B^2(s)},
\]
where $\mu$ is interpreted as the renormalization scale in Ref.[24]. This is incorrect($\mu$ is only a hard cutoff) . Due to the misuse of the cut off in the above integration($\mu$ is chosen to be $1 GeV^2$ in Ref.[24]), the value of two quark vacuum condensate is strongly underestimated in Ref.[24](see Table III and IV below). Here we stress the fact that one found from perturbative theory[26]
\[
B(p^2)=m_{0}\left(1-\frac{3\alpha_s}{4\pi}ln\left[\frac{p^2}{m^2_0}\right]+\cdot\cdot\cdot\right);
\]
i.e., the perturbative correction is proportional to the current quark mass $m_0$. Hence in a perturbative analysis $B(p^2)\propto m_0$, which vanishes in the chiral limit $m_0\rightarrow 0$. This means that the nonzero $B(p^2)$ even in the region of $p^2>1 GeV^2$ is an intrinsically nonperturbative effect in the chiral limit case. Therefore, the following contribution
\[
(-)\frac{3}{4\pi^2}\int^{\infty}_{\mu=1 GeV^2} ds \frac{sB(s)}{sA^2(s)+B(s)}
\]
should be taken into account in a correct treatment of the two quark vacuum condensate, that is, the upper limit of integration $\mu$ should be taken to be infinity.  

Another important vacuum condensate is the  four quark
condensate which are needed within the framework of QCD sum rule to describe properties of both mesons and baryons. From Eq.(20), it is easy to obtain
\begin{eqnarray}
&&\langle\tilde{0}|:\bar{q}(x)\Lambda^{(1)}q(x)\bar{q}(y)
\Lambda^{(2)}q(y):|\tilde{0}\rangle\nonumber\\
&&=-\left\{tr[\Sigma(y,x)\Lambda^{(1)}
\Sigma(x,y)\Lambda^{(2)}]
-tr[\Sigma(x,x)\Lambda^{(1)}]
tr[\Sigma(y,y)\Lambda^{(2)}]\right\}.
\end{eqnarray}
By means of Eq.(25), one can calculate all kinds of four quark condensates at the mean field level in GCM. For instance, in the case of
$\Lambda^{(1)}=\Lambda^{(2)}=\gamma_{5}$ and $x=y=0$, one finds
\begin{eqnarray}
&&\langle\tilde{0}|:\bar{q}(0)\gamma_{5} q(0)~\bar{q}(0)\gamma_{5} q(0):|\tilde{0}\rangle\nonumber\\
&&=(-12)\int\frac{d^4 p}{(2\pi^4)}\frac{d^4 q}{(2\pi^4)}\left\{
\left[\frac{B(p^2)}{A^2(p^2)p^2+B^2(p^2)}\right]
\left[\frac{B(q^2)}{A^2(q^2)q^2+B^2(q^2)}\right]\right.
\nonumber\\
&&\left.+
\left[\frac{A(p^2)}{A^2(p^2)p^2+B^2(p^2)}-C(p^2)\right]\left[\frac{A(q^2)}{A^2(q^2)q^2+B^2(q^2)}-C(q^2)\right]p\cdot q\right\}.
\end{eqnarray}
Notice, the integral in Eqs.(26) is free of UV divergence. However, if we follow the method proposed in Ref.[24], we have 
\begin{eqnarray}
&&\langle\tilde{0}|:\bar{q}(0)\gamma_{5} q(0)~\bar{q}(0)\gamma_{5} q(0):|\tilde{0}\rangle^{Meissner}\nonumber\\
&&=(-12)\int\frac{d^4 p}{(2\pi^4)}\frac{d^4 q}{(2\pi^4)}\left\{
\left[\frac{B(p^2)}{A^2(p^2)p^2+B^2(p^2)}\right]
\left[\frac{B(q^2)}{A^2(q^2)q^2+B^2(q^2)}\right]\right.\nonumber\\
&&\left.+
\left[\frac{A(p^2)}{A^2(p^2)p^2+B^2(p^2)}\right]\left[\frac{A(q^2)}{A^2(q^2)q^2+B^2(q^2)}
\right]p\cdot q\right\}\nonumber,
\end{eqnarray}
which is UV divergent. Here it should be noted that the calculation of vacuum condensates in the framework of GCM is free of UV divergence if one adopt the adequate subtraction mechanism. This result is also quite different from that of effective quark-quark interaction model, such as Nambu, Jona-Lasinio[5] and chiral constituent quark models[27].

The mixed quark gluon condensate $g\langle\bar{q}G_{\mu\nu}\sigma^{\mu\nu}q\rangle$ is also an important condensate within the framework of QCD sum rule. Now let us focus on the gluonic observables. Because the functional integration over the gluon field $A$ in Eq.(1) is quadratic for a given quark-quark interaction $D$ we can perform the integration over gluon fields analytically. Using the short hand notation for the integrations as in Ref.[24], we have
\[\int {\cal{D}} A~e^{-\frac{1}{2}A D^{-1} A +j A}\equiv e^{\frac{1}{2}j D j},~~~~\int {\cal{D}} A~A(x)e^{-\frac{1}{2}A D^{-1} A +j A}\equiv (jD)(x)e^{\frac{1}{2}j D j},\]
\begin{equation}
\int {\cal{D}} A~A(x)A(y)e^{-\frac{1}{2}A D^{-1} A +j A}
\equiv \left[D(x,y)+(jD)^2(x,y)\right]e^{\frac{1}{2}j D j}.
\end{equation}
This means that the gluon vacuum average renders effectively a quark color current $\bar{q}(x)\gamma_{\mu}\frac{\lambda^a_{C}}{2}q(x)$ together with the gluon two-point function $D$. In order to calculate the mixed quark gluon condensate
$g\langle \tilde{0}|:\bar{q}G_{\mu\nu}\sigma^{\mu\nu}q:|\tilde{0}\rangle$, we have to do integration over powers of gluon field $A$ and $A^2$. By means of Eq.(27) we find
\begin{eqnarray}
&&g\langle \tilde{0}|:\bar{q}(x)G_{\mu\nu}(x)\sigma^{\mu\nu}q(x):|
\tilde{0}\rangle\nonumber\\
&=&2g\langle \tilde{0}|:\bar{q}(x)\partial^{x}_{\mu}A^{a}_{\nu}(x)\frac{\lambda^{a}}{2}\sigma^{\mu\nu}q(x):|\tilde{0}\rangle+g^2f^{abc}\langle \tilde{0}|:\bar{q}(x)A^{b}_{\mu}(x)A^{c}_{\nu}(x)\frac{\lambda^a}{2}\sigma^{\mu\nu}q(x):|\tilde{0}\rangle\nonumber\\
&=&2ig^2\int d^4 z\left[\partial^{(x)}_{\mu} D(x,z)\right]
\langle \tilde{0}|:\bar{q}(x)\bar{q}(z)\gamma_{\nu}\frac{\lambda^a}{2} q(z)\frac{\lambda^a}{2}\sigma^{\mu\nu}q(x):|\tilde{0}\rangle \\
&-&f^{abc}\int d^4 z_{1} d^4 z_2 g^2D(x,z_1) g^2D(x,z_2)\langle \tilde{0}|:\bar{q}(x)\bar{q}(z_1)\gamma_{\nu}\frac{\lambda^b}{2} q(z_1)\bar{q}(z_2)
\gamma_{\nu}\frac{\lambda^c}{2}q(z_2)\frac{\lambda^a}{2}\sigma^{\mu\nu}q(x):|\tilde{0}\rangle\nonumber,
\end{eqnarray}
with $f^{abc}$ the structure constants of the SU(3) color group. As indicated by Eq.(28), in order to calculate the mixed quark gluon condensate, one must calculate the four quark and six quark condensates in advance. Applying Eqs.(20) and (21), we have
\begin{eqnarray}
&&g\langle \tilde{0}|:\bar{q}(x)G_{\mu\nu}(x)\sigma^{\mu\nu}q(x):|
\tilde{0}\rangle=-8ig^2\int d^4 z\left[\partial^{(x)}_{\mu}D(x,z)\right]tr_{D}
\left[\Sigma(z,x)\sigma_{\mu\nu}\Sigma(x,z)\gamma_{\nu}\right]\nonumber\\
&&+12i\int d^4 z_1 d^4 z_2 g^2D(x,z_1) g^2D(x,z_2)tr_{D}\left[\Sigma(z_2,x)\sigma_{\mu\nu}\sigma(x,z_2)\gamma_{\mu}\Sigma(z_1,z_2)\gamma_{\nu}\right]\nonumber\\
&&=-36\int\frac{d^4 p}{(2\pi)^4}\frac{B(p^2)[2-A(p^2)]p^2}{A^2(p^2)p^2+B^2(p^2)}-\frac{81}{4}\int\frac{d^4 p}{(2\pi)^4}\frac{B(p^2)\left[B^2(p^2)+2p^2\left(A(p^2)-1\right)A(p^2)\right]}{A^2(p^2)p^2+B^2(p^2)}\nonumber\\
&&-12\int\frac{d^4 p}{(2\pi)^4}\int\frac{d^4 q}{(2\pi)^4}g^2D(p-q)
\left[\frac{16-9A(p^2)}{A^2(p^2)p^2+B^2(p^2)}+9C(p^2)\right]
B(p^2)p\cdot q C(q^2)\nonumber\\
&&+\frac{9}{2}\int \frac{d^4 p}{(2\pi)^4}\left[9A(p^2)-1\right]B(p^2)C(p^2)p^2,
\end{eqnarray}
where $tr_{D}$ denotes that the trace in Eq.(29) is to be taken in Dirac space only, whereas the color trace has been separated out. It is easy to find that our result for the mixed quark gluon condensate is substantially different from that in Ref.[24](see Eq.(25) in Ref.[24]). The result for the mixed quark gluon condensate in Ref.[24] keeps only the contribution of the first and the second term of the right hand in Eq.(29). 

By means of Eq.(22), Eq.(29) can be further reduced to:
\begin{eqnarray}
&&g\langle \tilde{0}|:\bar{q}(x)G_{\mu\nu}(x)\sigma^{\mu\nu}q(x):|
\tilde{0}\rangle=
\frac{81}{2}\int\frac{d^4 p}{(2\pi)^4}B(p^2)\left[\frac{8}{9}p^2C(p^2)+A(p^2)p^2C(p^2)-\frac{3}{2}\right]\nonumber\\
&&+\frac{9}{4}\int\frac{d^4 p}{(2\pi)^4}\frac{B(p^2)}{A^2(p^2)p^2+B^2(p^2)}\left\{\frac{18A(p^2)-32}{C(p^2)}-A(p^2)p^2\left[9A(p^2)-16\right]\right\}\nonumber\\
&&=\frac{81}{32\pi^2}\int^{\infty}_{0}sds~B(s)\left[\frac{8}{9}sC(s)+A(s)sC(s)-\frac{3}{2}\right]\nonumber\\
&&+\frac{9}{64\pi^2}\int^{\infty}_{0}sds~\frac{B(s)}{A^2(s)s+B^2(s)}\left\{\frac{18A(s)-32}{C(s)}-A(s)s\left[9A(s)-16\right]\right\}.
\end{eqnarray}
\begin{center}
\begin{table}
\caption{The two quark and mixed quark gluon condensate for $g^2D^{(1)}_{IR}(q^2)$=$3\pi^2 \frac{\chi^2}{\Delta^2}e^{\frac{q^2}{\Delta}}$}
\begin{tabular}{cccc} 
\hline\hline            
$\Delta[GeV^4]$~~~~ &$\chi[GeV]$~~~~ & (--)$\langle\tilde{0}| :\bar{q}q:|\tilde{0}\rangle^\frac{1}{3}$~(MeV)& (--)
$g\langle\bar{q}G_{\mu\nu}\sigma^{\mu\nu}q\rangle^{\frac{1}{5}}~(MeV)$ \\ \hline
0.200  &1.65     & 252                  & 296  \\
0.020  &1.55     & 212                  & 353 \\
0.002  &1.45     & 190                  & 349\\ \hline\hline
\end{tabular}
\end{table}
\end{center}
\begin{center}
\begin{table}
\caption{The two quark and mixed quark gluon condensate for $g^2D^{(2)}_{IR}(q^2)$=$4\pi^2 d\frac{\chi^2}{q^4+\Delta}$}
\begin{tabular}{cccc} 
\hline\hline            
$\Delta[GeV^4]$~~~~ &$\chi[GeV]$ ~~~~ & (--)$\langle\tilde{0}| :\bar{q}q:|\tilde{0}\rangle^\frac{1}{3}$~(MeV)& (--)
$g\langle\bar{q}G_{\mu\nu}\sigma^{\mu\nu}q\rangle^{\frac{1}{5}}~(MeV)$ \\ \hline
$10^{-1}$  &1.83  & 344                     & 301  \\
$10^{-4}$  &1.02  & 264                     & 336 \\
$10^{-7}$  &0.83  & 264                     & 393\\ \hline\hline
\end{tabular}
\end{table}
\end{center}

\vspace*{-1.2 cm}

From Eqs.(24) and (30), we can calculate the two quark condensate and the mixed quark gluon condensate. In Table I and II, we display the numerical results of the two quark and the mixed quark gluon condensate for three different sets of parameters of two different model gluon propagators(Eq.(12)and Eq.(13)).

Table I and II show that the result of the two quark condensate and the mixed quark gluon condensate is compatible with the corresponding ``standard'' value determined from QCD sum rule[1,2]. Here we want to stress that the form of gluon propagator and the corresponding parameters are somewhat flexible, as indicated by the authors in Ref.[28]. Thus our result about the two quark and the mixed quark gluon condensate may provide a further constraint on the the form and parameters of the gluon propagator.

It should be noted that the cut-off $\mu$ plays an essential role in calculating the value 
of the condensates in Refs.[24,25]. However, as is shown above, in a correct treatment of vacuum condensate, the cut off $\mu$ should be taken to be infinity. This situation is very similar to the determination of vacuum condensates in the instanton liquid model where the upper 
limit of integration $\mu$ is taken to be infinity[29,30]. Because the calculation 
is numerical, one has to use a very large but finite value of upper limit of integration. In our numerical calculation, the upper limit is chosen to be $500~GeV^2$, which is large enough to ensure that the calculated value is independent of the choice of the upper limit of integration. This result can be seen from Table III and IV.

\begin{center}
\begin{table}
\caption{The sensitivity of the vacuum condensates to the choice of the upper limit of integration $\mu$ in Eqs.(24) and (30) for $g^2D^{(1)}_{IR}(q^2)$=$3\pi^2 \frac{\chi^2}{\Delta^2}e^{\frac{q^2}{\Delta}}$($\Delta=0.020GeV^4$,~~$\chi=1.55GeV$)}
\begin{tabular}{ccc} 
\hline\hline           
$\mu$~($GeV^2$)~~~~~~~~~~~ & (--)$\langle\tilde{0}| :\bar{q}q:|\tilde{0}\rangle^\frac{1}{3}~(MeV)$~~~~~~~~~~
& (--)$g\langle\bar{q}G_{\mu\nu}\sigma^{\mu\nu}q\rangle^{\frac{1}{5}}~(MeV)$ \\ \hline
1        & 170                     &  352 \\
100      & 205                     &  353\\
300      & 210                     &  353\\
450      & 212                     &  353\\ 
500      & 212                     &  353 \\ \hline\hline
\end{tabular}
\end{table}
\end{center}
\begin{center}
\begin{table}
\caption{The sensitivity of the vacuum condensates to the choice of the upper limit of integration $\mu$ in Eqs.(24) and (30) for $g^2D^{(2)}_{IR}(q^2)$=$4\pi^2 d\frac{\chi^2}{q^4+\Delta}$($\Delta=10^{-4}GeV^4$,~~$\chi=1.02GeV$)}
\begin{tabular}{ccc} 
\hline\hline           
$\mu$~($GeV^2$)~~~~~~ & (--)$\langle\tilde{0}| :\bar{q}q:|\tilde{0}\rangle^\frac{1}{3}~(MeV)$~~~~~~~& (--)
$g\langle\bar{q}G_{\mu\nu}\sigma^{\mu\nu}q\rangle^{\frac{1}{5}}~(MeV)$ \\ \hline
1        & 177                  & 333  \\
100      & 254                  & 336  \\
300      & 262                  & 336  \\
450      & 264                  & 336  \\ 
500      & 264                  & 336  \\ \hline\hline
\end{tabular}
\end{table}
\end{center}

\vspace*{-2.6 cm}

\begin{center}
{\bf \large III. DISCUSSION}
\end{center}

To summarize, in the present paper, we provided a general recipe to calculate the vacuum condensates at the mean field level in the framework of GCM. This approach is quite different from that in previous studies[24,25] in the following aspects: 

\begin{description}
\item{~ 1.} The first difference between the approach proposed in Refs.[24,25] and our's comes from the definition of vacuum condensate. The definition of the vacuum condensate in Refs.[24,25] is only the corresponding Green function taken at one point in coordinate space(see Eq.(19) of Ref.[24])), which is not the expectation value of the ``normal'' product over the vacuum $|\tilde{0}\rangle$. Therefore, it is not consistent with that in the QCD sum rule approach and the spurious contribution of the perturbative terms have not been subtracted. In the calculation of the two quark condensate, the contribution of the subtraction term vanishes which explains why our formula for the two quark condensate exactly coincides with the corresponding ones in Ref.[24](in this case, the main numerical difference between the two parties comes from the integration region between $\mu$=1 $GeV^2$ to infinity). In other cases, the contribution of the corresponding subtraction terms do not vanish, and this will cause the substantial difference between the expression of the vacuum condensates in Ref.[24] and in ours(for example, see the expression of the mixed quark gluon condensate in Eq.(29)). In the present paper, an adequate subtraction mechanism which is consistent with the definition of vacuum condensate(the vacuum expectation value of the ``normal'' product of fields) in QCD sum rule is adopted to calculate vacuum condensates(employing Wick theorem and considering in addition the usual Nambu-Goldstone phase a perturbative Wigner phase of GCM). Within this approach all kinds of vacuum condensates are free of UV divergence. This is reasonable and what is to be expected. In addition, due to the fact that the definition of the vacuum condensate in Ref.[24] is not the vacuum expection value of the normal product of fields, it will be plagued with UV divergence in some cases(see Eq.(26) and related items). 
\end{description}

\begin{description}
\item{~ 2.} The second disagreement between the two parties is the value of the upper limit of integration $\mu$. In the calculation of vacuum condensates of the present paper, all upper limit of integration is taken to be $\infty$. This is very similar to the determination of vacuum condensates in the instanton liquid model[29,30], which is different from that in Refs.[24] and [25]. Because of the misuse of cut off $\mu=1 GeV^2$ in Ref.[24], the value of the two quark condensate is strongly underestimated there. In addition, if we use Eq.(25) in Ref.[24] and let the cutoff $\mu$ to be $\infty$ to calculate the mixed quark gluon condensate, we get a value which is several times larger than the ``standard'' value of the mixed quark gluon condensate obtained by QCD sum rule.
\end{description}

The numerical results of the two quark condensate $\langle\bar{q}q\rangle$ and the mixed quark gluon condensate $g\langle\bar{q}G_{\mu\nu}\sigma^{\mu\nu}q\rangle$ obtained by the modified approach show that they are compatible with the values obtained from QCD sum rule. These results can be used as a consistency check of the GCM and provide a reference for further study. Finally, we want to stress that 
the GCM is not renormalizable. Therefore, the scale at which a condensate is defined in our approach is a typical hadronic scale, which is implicitly determined by the model-gluon propagator $g^2D(q^2)$ and the solution of the rainbow DS equation(10-11). This situation is very similar to the determination of vacuum condensates in the instanton liquid model where the scale is set by the inverse instanton size[29,30].

\vspace*{0.2 cm}
\noindent{\large \bf Acknowledgments}

This work was supported in part by the National Natural Science Foundation
of China under Grant Nos 19975062, 10175033 and 10135030.

\vspace*{0.4 cm}
\noindent{\large \bf References}
\begin{description}
\item{[1]} M. Shifman, A. Vainshtein and V. Zakharov, Nucl. Phys.
{\bf B147}, 385 (1979).
\item{[2]} L. Reinders, H. Rubinstein and S. Yazaki, Phys. Rep.
{\bf 127}, 1 (1985); S. Narison, QCD Spectral Sum Rules (World Scientific, Singapore
, 1989), and references therein.
\item{[3]} G. E. Brown and M. Rho, Phys. Rev. Lett. {\bf 66}, 2720 (1991).
\item{[4]} T. D. Cohen, R. J. Furnstahl, and D. K. Griegel, Phys. Rev. {\bf C45}, 1881 (1992).
\item{[5]} S. P. Klevansky, Rev. Mod. Phys. {\bf 64}, 649 (1992).
\item{[6]} T. Renk, R. A. Schneider and W. Weise, Nucl. Phys. {\bf A699}, 1c (2002).
\item{[7]} A. Bender, W. Detmold, A. W. Thomas, Phys. Lett. {\bf B516}, 54 (2001).
\item{[8]} C. D. Roberts and S. M. Schmidt, Prog. Part. Nucl. Phys. {\bf 45}, 1 (2000), and references therein.
\item{[9]} R. T. Cahill and C. D. Roberts, Phys. Rev. {\bf D32}, 2419 (1985); P. C. Tandy, Prog. Part. Nucl. Phys. 39, 117 (1997); R. T. Cahill and S. M. Gunner, Fiz. {\bf B7}, 17 (1998), and references therein.
\item{[10]} C. D. Roberts and A. G. Williams, Prog. Part. Nucl. Phys. {\bf 33}, 477 (1994), and references therein.
\item{[11]} M. R. Frank and T. Meissner, Phys. Rev {\bf C57}, 345 (1998).
\item{[12]} C. D. Roberts, R. T. Cahill, and J. Praschiflca, Ann. Phys. (N. Y.){\bf 188}, 20 (1988).
\item{[13]} P. Maris, C. D. Roberts, and P. C. Tandy, Phys. Lett {\bf B420}, 267 (1998).
\item{[14]} C. D. Roberts, R. T. Cahill, M. E. Sevior, and N. Ianella, Phys. Rev. {\bf D49}, 125 (1994). 
\item{[15]} M. R. Frank and T. Meissner, Phys. Rev {\bf C53}, 2410 (1996).
\item{[16]} R. T. Cahill and S. Gunner, Phys. Lett. {\bf B359}, 281 (1995); Mod. Phys. Lett. {\bf A10}, 3051 (1995).
\item{[17]} P. Maris and C. D. Roberts, Phys. Rev {\bf C56}, 3369 (1997).
\item{[18]} Xiao-fu L\"{u}, Yu-xin Liu, Hong-shi Zong and En-guang Zhao, Phys. Rev. {\bf C58}, 1195 (1998);  Hong-shi Zong, Yu-xin Liu,  Xiao-fu L\"{u}, Fan Wang, and En-guang Zhao, Commun Theor. Phys. (Beijing, China) {\bf 36}, 187 (2001); Hong-shi Zong, Xiang-song Chen, Fan Wang, Chao-hsi Chang and En-guang Zhao,Phys. Rev. {\bf C66}, 015201 (2002).
\item{[19]} C. Burden, C. D. Roberts, and M. Thomson, Phys. Lett. {\bf B371}, 163 (1996).
\item{[20]} C. Burden and D. Liu, Phys. Rev. {\bf D55}, 367 (1997); M. A. Ivanov, Yu. L. Kalinovskii, P. Maris, and C. D. Roberts, Phys. Lett. {\bf B416}, 29 (1998); Phys. Rev. {\bf C57}, 1991 (1998). 
\item{[21]} A. Bender, D. Blaschke, Y. Kalinovskii, and C. D. Roberts, Phys. Rev. Lett. {\bf 77}, 3724 (1996).
\item{[22]} M. R. Frank, P. C. Tandy, and G. Fai, Phys. Rev {\bf C43}, 2808 (1991); M. R. Frank and P. C. Tandy, Phys. Rev {\bf C46}, 338 (1992); C. W. Johnson, G. Fai, and M. R. Frank, Phys. Lett. {\bf 386}, 75 (1996).
\item{[23]} R. T. Cahill, Nucl. Phys. {\bf A543}, 63 (1992).
\item{[24]} T. Meissner, Phys. Lett. {\bf B405}, 8 (1997).
\item{[25]} Hong-shi Zong, Xiao-fu L\"{u}, Jian-zhong Gu, Chao-hsi Chang and En-guang Zhao,Phys. Rev. {\bf C60}, 055208 (1999); Hong-shi Zong, Xiao-fu L\"{u}, Fan Wang, Chao-hsi Chang, and En-guang Zhao, Commun Theor. Phys. (Beijing, China) {\bf 34}, 563 (2000).
\item{[26]} R. Pascual, and R. Tarrach, (1984) pp.67-70
: Renormalisation for the Practitioner. Springer-Verlag, Berlin.
\item{[27]} W. Broniowski, M. Polyakov, Hyun-chi Kim, and K. Goeke, Phys. Lett. {\bf B438}, 
242 (1998).
\item{[28]} C. D. Roberts, A. G. Williams, and G. Krein, Intern. Journal Mod. Phys. {\bf A7}, 5607 (1992).
\item{[29]} M. V. Polyakov and C. Weiss, Phys. Lett. {\bf B387}, 841 (1996).
\item{[30]} A. E. Dorokhov, S. V. Esaibegian, and S. V. Mikhailov, Phys. Rev. {\bf D56}, 4062 (1997).
\end{description}

\end{document}